\documentclass[12pt]{article}

\usepackage{amsmath,amssymb,amsfonts,amsbsy}
\usepackage{cite}
\usepackage{graphicx}
\usepackage{wrapfig}
\usepackage{epsfig}
\usepackage{multicol}

\begin{document}

\setcounter{section}{0}
\setcounter{subsection}{0}
\setcounter{equation}{0}
\setcounter{figure}{0}
\setcounter{footnote}{0}
\setcounter{table}{0}


\begin{center}
\textbf{\large STUDY OF THE DP-ELASTIC SCATTERING AT 2 GEV}

\vspace{0cm}

\underline {A.A.~Terekhin}$^{*\dag}$, Yu.V.~Gurchin$^{*}$, A.Yu.~Isupov$^{*}$, A.N.~Khrenov$^{*}$,\\
A.K.~Kurilkin$^{*}$, P.K.~Kurilkin$^{*}$, V.P.~Ladygin$^{*}$, N.B.~Ladygina$^{*}$, \\
 S.M.~Piyadin$^{*}$, S.G.~Reznikov$^{*}$, I.E.~Vnukov$^{**}$.

\vspace{1mm}
$^{*}${\em LHEP-JINR, 141980, Dubna, Moscow region, Russia}\\
$^{**}$ {\em BelSU, 308015,Belgorod,Russia}

 $\dag$ \emph{E-mail: aterekhin@jinr.ru}
\end{center}

\begin{abstract}
The results on the measurements of $dp$-elastic scattering cross section at the energy 2 GeV at Internal Target Station at the  Nuclotron JINR are reported. The data were obtained for the angular range of $70^o-107^o$ in the c.m.s. by using CH$_2$ and C targets. The results are compared with the existing data and with the theoretical calculations based on the relativistic multiple scattering theory. 
 
\end{abstract}


\textbf{\Large Introduction}

\vskip 5mm

\par The dp-elastic scattering reaction is the longtime  subject of the theoretical and experimental  investigations. Many experiments were performed by using the different targets and the accelerated particles beams. Today, there are quite extensive experimental material. Also now the different theoretical models  are developed to describe of data at the different  energies: the Faddeev calculations in the momentum space \cite{Witala} and configuration space \cite{Friar}, and variational calculations based on the solution of the three-particle Schroedinger equation  \cite{Kievsky,Viviani,Deltuva}.The momentum- space  Faddeev equations for three-nucleon scattering can now be solved with high accuracy for the most modern two- and three- nucleon forces below 200 MeV/n of the projectile energy \cite{Glockle,Kuros}. The discrepancy between the theory and experiment is increasing with the energy increasing indicating the possibility of relativistic effects. The theoretical calculations using not only 2N forces but also different 3N forces \cite{Coon,Rudliner} give the best agreement with experimental data. 
\par The experimental material for $dp$-elastic scattering covers  the energy range from tens to thousands MeV/n. The precise data were obtained at RIKEN at the energies of 70, 100 and 135 MeV/n \cite{Sekiguchi} for the angular range of $10^\circ < \theta^* < 180^\circ$. The analogous experiment was performed in RCNP at the energy of 250 MeV/n \cite{Hatanaka}, where the data on the cross section and complete set of proton spin observables were obtained.  The goal of these experiments was to study the 3NF contribution and to test modern models of the three-nucleon forces. 
\par The differential cross section data at energies 470 and 590 MeV/n in the backward hemisphere were obtained at the National Aeronautics and Administration Space  Radiation Effects Laboratory \cite{Alder}.  The absolute differential cross section was measured  at 641.3 and 792.7 MeV/n  in the angular range of $35^\circ-115^\circ$ and $35^\circ-140^\circ$, respectively \cite{Culmez}. The results can be fit with a relativistic multiple-scattering theory which uses off-mass-shell extrapolations of the nucleon-nucleon amplitudes suggested by the structure of derivative meson-nucleon couplings \cite{Bleszynski}. Relativistic-impulse-approximation calculations do not describe these data \cite{Shepard}. The data were obtained at the Brookhaven National Laboratory (BNL) at 1 GeV/n for the angles of  $10^\circ< \theta^* <170^\circ$ \cite{Bennet}. The new data on the differential cross section of the $dp$-elastic scattering at 1.25 GeV/n were obtained with HADES detector \cite{Kurilkin}. The experimental data are described by the relativistic  multiple scattering theory which takes both single and double interactions into account \cite{Ladygina}. 
\par The transition to higher energies (few hundred MeV/n and higher) will allow one to understand the mechanism of manifestation of the fundamental degrees of freedom at distances of the order of the nucleon size. Glauber scattering theory which takes both single and double interactions in this case is a classic approach \cite{Franco,Franco1}.
\par The purpose of DSS (Deuteron Spin Structure) project \cite{Ladygin} is the broadening of the energy and angular ranges of measurement of different observables in processes including 3-nucleon systems. The experiments to study of the $dp$-elastic scattering at Internal  Target Station (ITS) \cite{Malakhov} at the Nuclotron in the range from 150 to 1000 MeV/n are performed in the frame of this project. The experimental setup allows one to obtain the different observables from 60$^o$ to 140$^o$ in the c.m.s. The new preliminary differential cross section data were obtained at ITS  Nuclotron at the energies from 250 to 440 MeV/n \cite{Gurchin}. Recently, the deuteron vector and tensor analyzing powers have been obtained at 440 MeV/n \cite{KurilkinAyy}. The angular dependence of the differential cross section obtained at 2GeV is presented in this paper. The results are compared with the existing data \cite{Terekhin} and \cite{Bennet} and with the theoretical calculations based on the relativistic multiple scattering theory \cite{Ladygina}.
 \section{Experiment}
\par The measurements were performed at ITS \cite{Malakhov} at the Nuclotron JINR by using 10$\mu m$ CH$_2$ and 8$\mu m$ C targets. New ITS DAQ system was used during the data taking \cite{Isupov}. The elastically-scattered deuterons and protons were counted by two pairs of detectors placed symmetrically with respect to the beam direction. This allows one to improve the quality of the experiment. During the methodical measurements all deuteron- and proton-counters (the description of which can be found in \cite{Gurchin85}) are based on the FEU-85. During the main measurements these detectors were replaced to the counters based on the Hamamatsu H7416MOD previously used in the experiment \cite{KurilkinAyy}. Another two detectors (PP-detectors) based on the  FEU-85 and FEU-63 were used to count the quasi-elastically scattered protons. Each  of such a counter consists of the $\Delta E$ and $E$ - detectors \cite{Piyadin}. The layout of the counters with respect to the beam direction is shown in Fig.1. The D$_{1,2}$, P$_{1,2}$ and  PP$_{1,2}$ are deuteron-, proton- and PP-detectors, respectively. All the counters were placed in horizontal plane. The DP-detectors were rotated to give an angular range of the  $\theta_{lab} = 19^o$ to $50^o$ ($\theta_{c.m.} = 70^o$ to $120^o$). The  precision of  the detectors mount is $0.3^o$ in the laboratory system, which corresponds to $~0.6^o$ in the c.m.s. The PP-detectors were mounted at the angle corresponding to quasi-elastic scattering at $\theta_{c.m.} = 90^o$ and remained stationary throughout the experiment. These detectors were used as relative luminosity monitors. The sizes of the P-,D- and PP- counters are 20x60x20 mm$^3$, 50x50x20 mm$^3$ and $\phi$100x200 mm$^3$, respectively. The distances from proton-, deuteron- and monitor- counters to the point of the beam interaction with the target are 60, 56 and 100 cm, respectively. The angular spans of P-, D- and PP- detectors were $2^o$, $5^o$, $10^o$ in the laboratory system, which corresponds to $~4^o$, $~10^o$ and $~20^o$ in the c.m.s., respectively. The characteristics of the detectors are shown in Table 1.
\begin{figure}[h]
  \begin{center}
    \includegraphics[width=130mm]{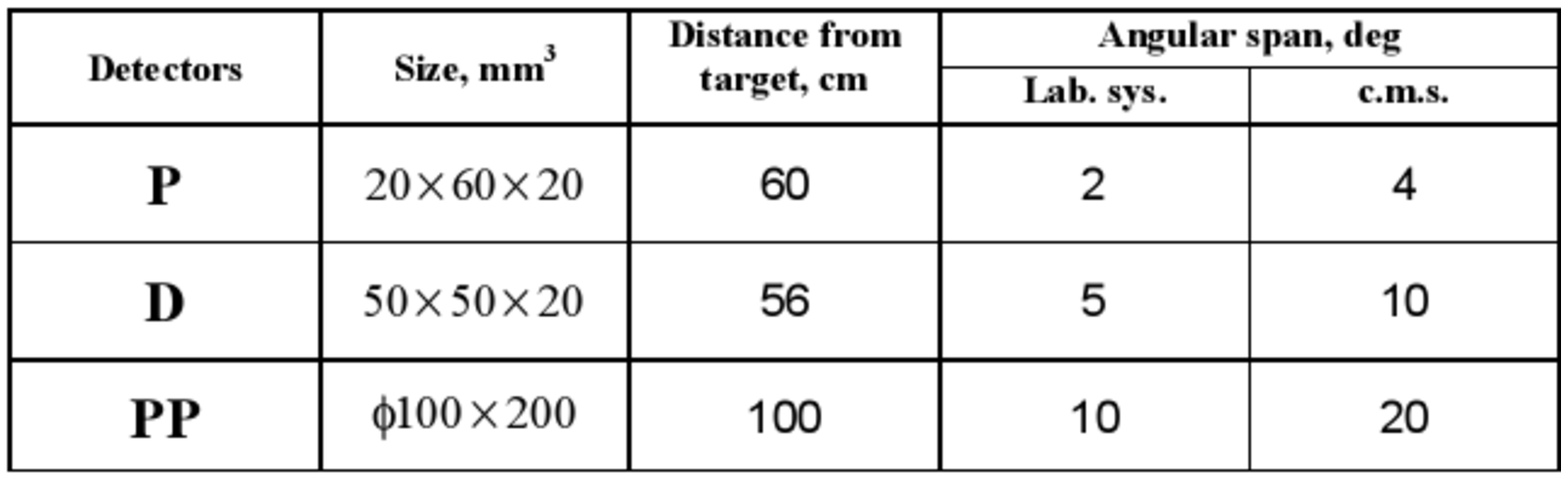}
  \end{center}
  \begin{center}
  Table 1. \footnotesize The characteristics of the detectors.
  \end{center}
\label{fig:label}
\end{figure} 

\par The VME based data acquisition system was used for the data taking from scintillation detectors. TQDC16 module \cite{nica}  allows one to measure the amplitude and time appearance of the signal simultaneously. Each module is separated into two parts with 8 input channels  having own first level trigger logics. In the current experiment the first level trigger signal was appeared when the signal from one module part coincides with the signal from any channel of  other part. 
\par The methodical measurements by using the scintillation counters based on the FEU-85 were performed. The DP- and monitor- detectors were mounted at angles $\theta_{c.m.} =75^o$ and $\theta_{c.m.} =90^o$ in the c.m.s., respectively. The analysis of subtraction of the time signal taken from the deuteron- and proton- detectors showed that the $dp$-elastic scattering events and background cannot be selected \cite{TerekhinPos}. Therefore, the scintillation counters based on the Hamamatsu were used for the current measurements. 
\par The results of measurements are shown in Fig.2. One can see that the use of Hamamatsu photomultipliers allow selecting the $dp$-elastic scattering events and background from carbon.
\begin{figure}[h]
\begin{multicols}{2}
\hfill
\includegraphics[width=6cm]{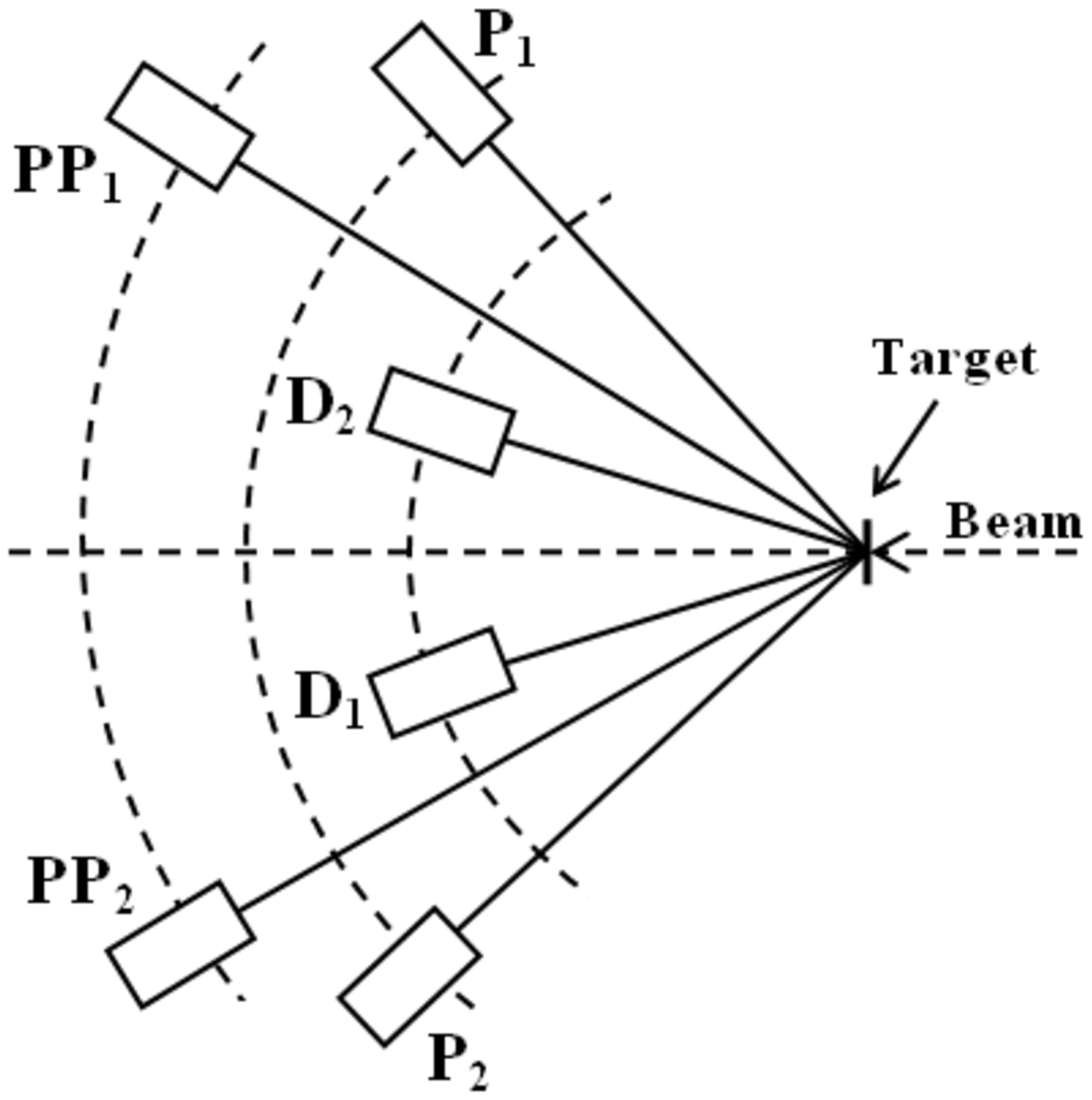}
\hfill
\caption{\footnotesize Layout of the counters with respect to the beam direction.D$_{1,2}$, P$_{1,2}$- deuteron and proton detectors, $PP_{1,2}$- the relative luminosity monitors.}
\label{1}
\hfill
\includegraphics[width=8cm]{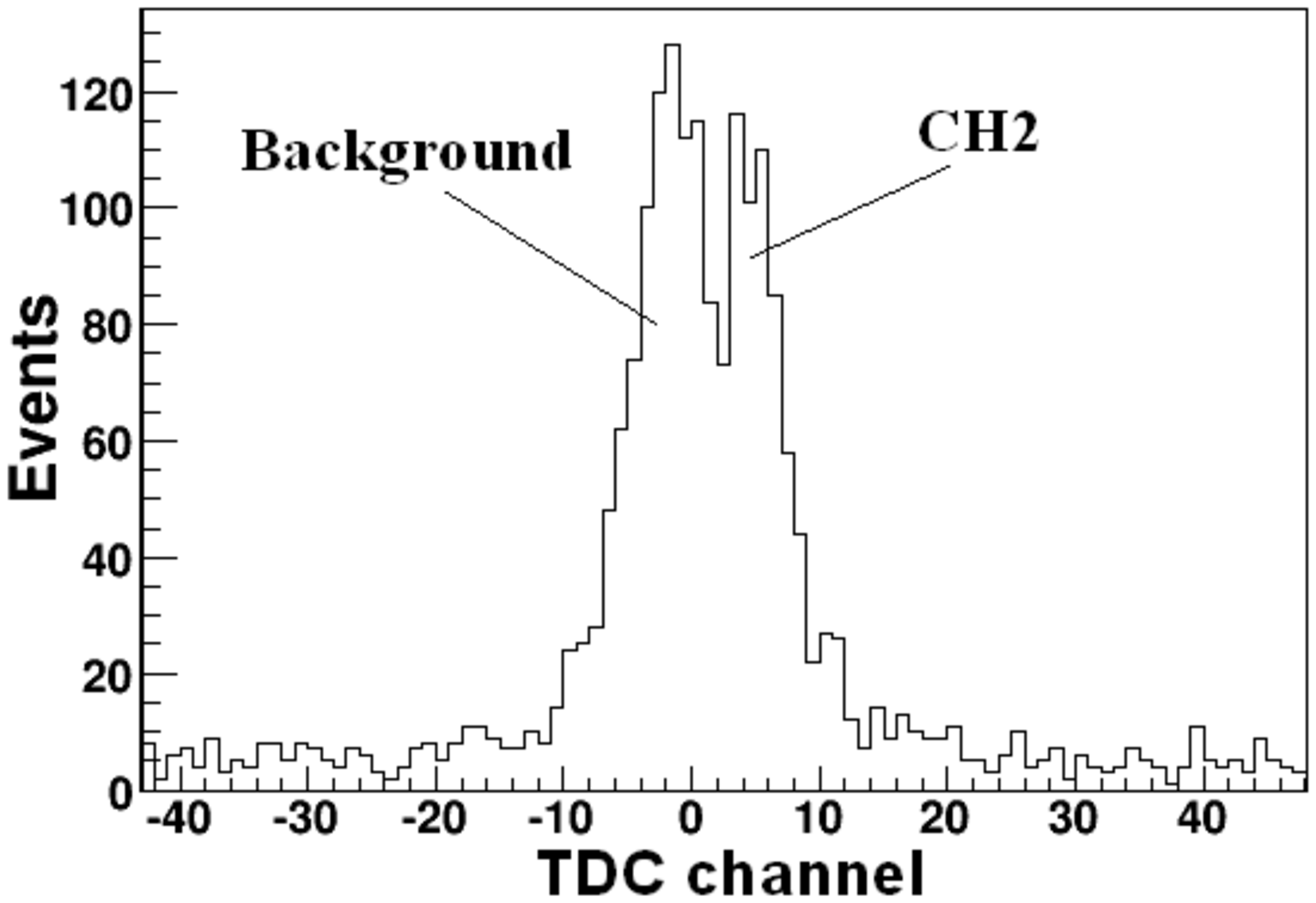}
\hfill
\caption{\footnotesize The subtraction of the timing signal from deuteron- and proton- counters for $\theta_{c.m.} = 70^o$ in the c.m.s.. The data were obtained by using the counters based on the Hamamatsu H7416MOD.}
\label{2}
\end{multicols}
\end{figure} 

 \section{Data analysis }
\par The data analysis was performed in the following way. Firstly, the identity of detectors work of each pair was tested. The target is moved by using a stepping motor. The correlation of  motor pulse  and time appearance of triggers are shown in Fig.3. The smooth line corresponds to the time when the target is located in the beam. The time range  after 1800 msec corresponds to the case when  the target is removed from the beam. The acquisition time of triggers  has been divided into the consecutive interval. The value of the each interval equal to 200 ms. The reconstructed events ("true" triggers) for DP-detectors are defined as the coincidence of signals from the one proton-counter with  the corresponding D-counter. For monitor counters the reconstructed events are defined as the coincidence of  both $\Delta E$ and $E$ detectors. The ratio of signal coincidences of D$_1$- and P$_1$- counters to signal coincidences of D$_2$- and P$_2$- counters was made for the each interval. The dependence of the value $R = N_1/N_2$ from the time interval, where $N_{1(2)}$ - the 
number of coincidences of P$_{1(2)}$- and D$_{1(2)}$- detectors counts, is shown in Fig. 4. One can see, that the value of the ratio is $\approx$1 in the domain when the target is within the beam. The similar results were obtained for the PP$_{1(2)}$ detectors. This demonstrates small geometrical	misalignment of the experimental setup with respect to the beam direction. 

\begin{figure}[h]
\begin{multicols}{2}
\hfill
\includegraphics[width=75mm]{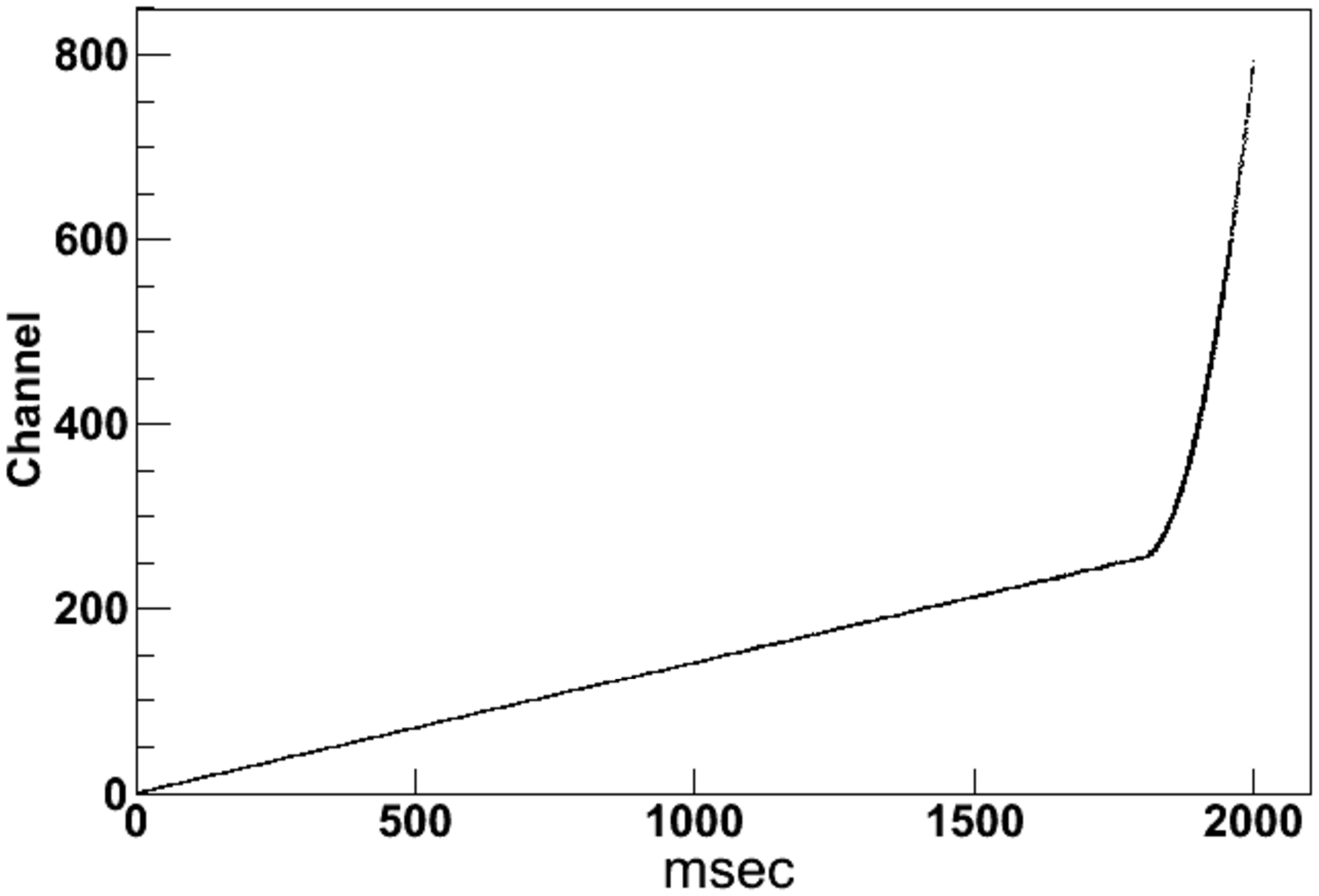}
\hfill
\caption{\footnotesize The correlation of motor pulse and time appearance of triggers.
 }
\label{3}
\hfill
\includegraphics[width=75mm]{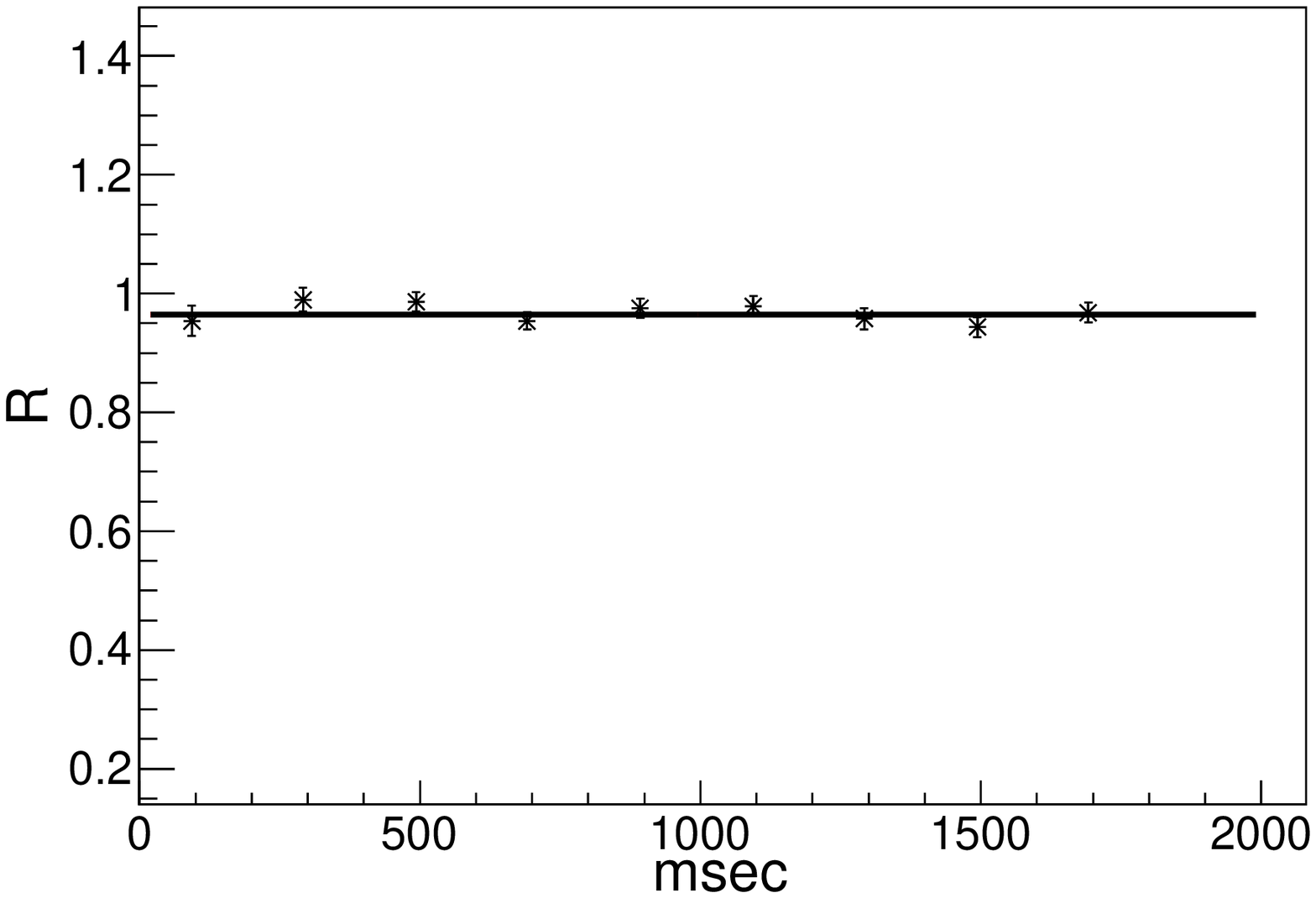}
\hfill
\caption{\footnotesize The ratio of signal coincidences of D$_1$- and P$_1$- counters to signal coincidences of D$_2$- and P$_2$- counters for $\theta_{c.m.} = 100^o$ in the c.m.s. as a function of the acquisition time of triggers.}
\label{4}
\end{multicols}
\end{figure} 
\par The procedure to obtain differential cross section data  was made by analysis of the amplitude spectra. The graphical cut was imposed on the signal amplitudes correlation for  D- and P- detectors to select $dp$-elastic scattering particles (Fig. 5). 
\par The estimation of the background in the amplitude data was performed by using the  temporary gates on the deuteron and proton time difference spectra. The subtraction of the timing signal from deuteron- and proton- counters was made by using the cut for signal amplitudes correlation (Fig 6). In this distribution the $dp$-elastic scattering events (I domain) and the background (II and III domains) are selected so that the width of both domains are equal.
\begin{figure}[h]
\begin{multicols}{2}
\hfill
\includegraphics[width=65mm]{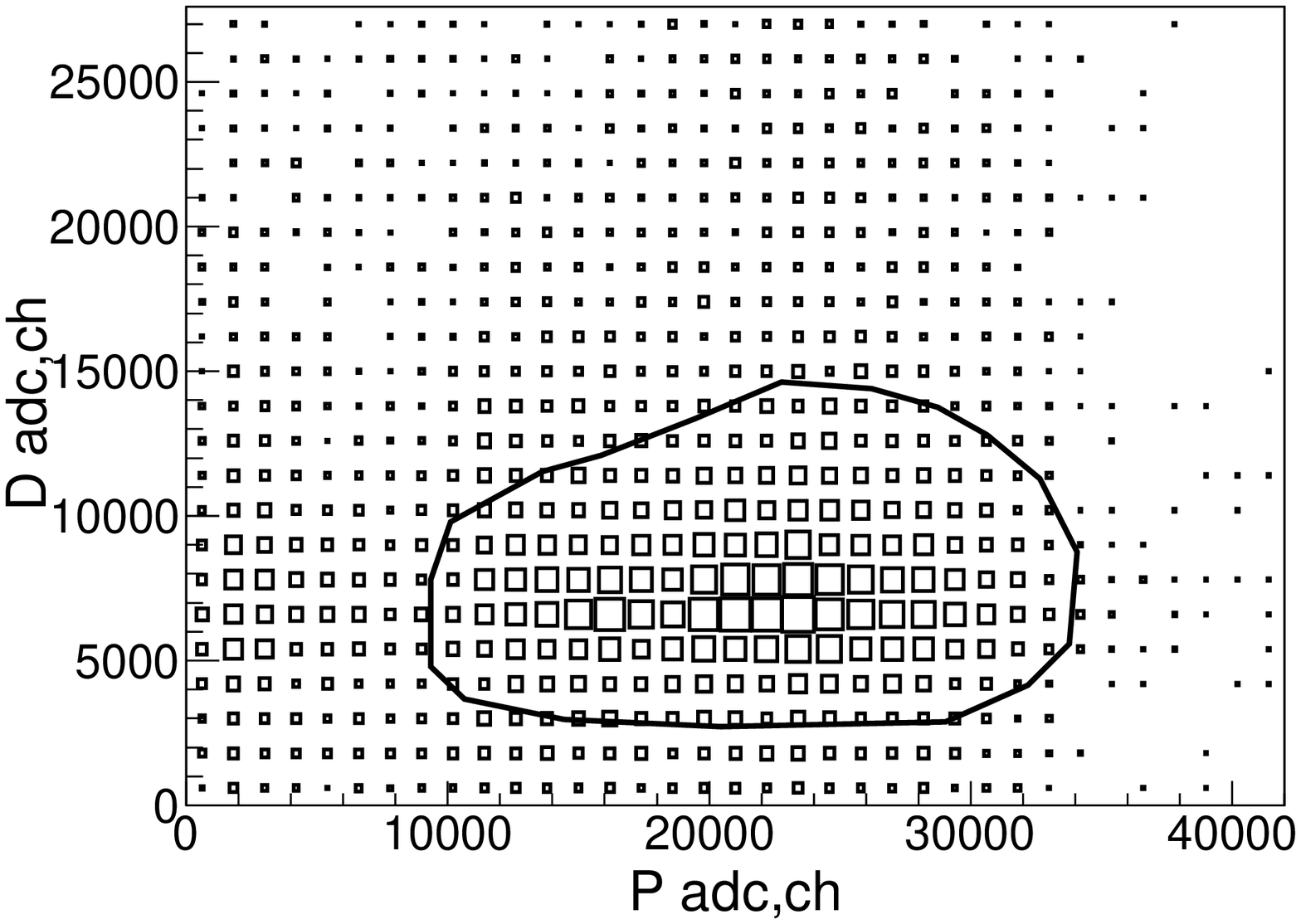}
\hfill
\caption{\footnotesize The signal amplitudes correlation for  deuteron- and proton detectors. The solid line is the graphical cut to select $dp$-elastic scattering events.
 }
\label{2}
\hfill
\includegraphics[width=71mm]{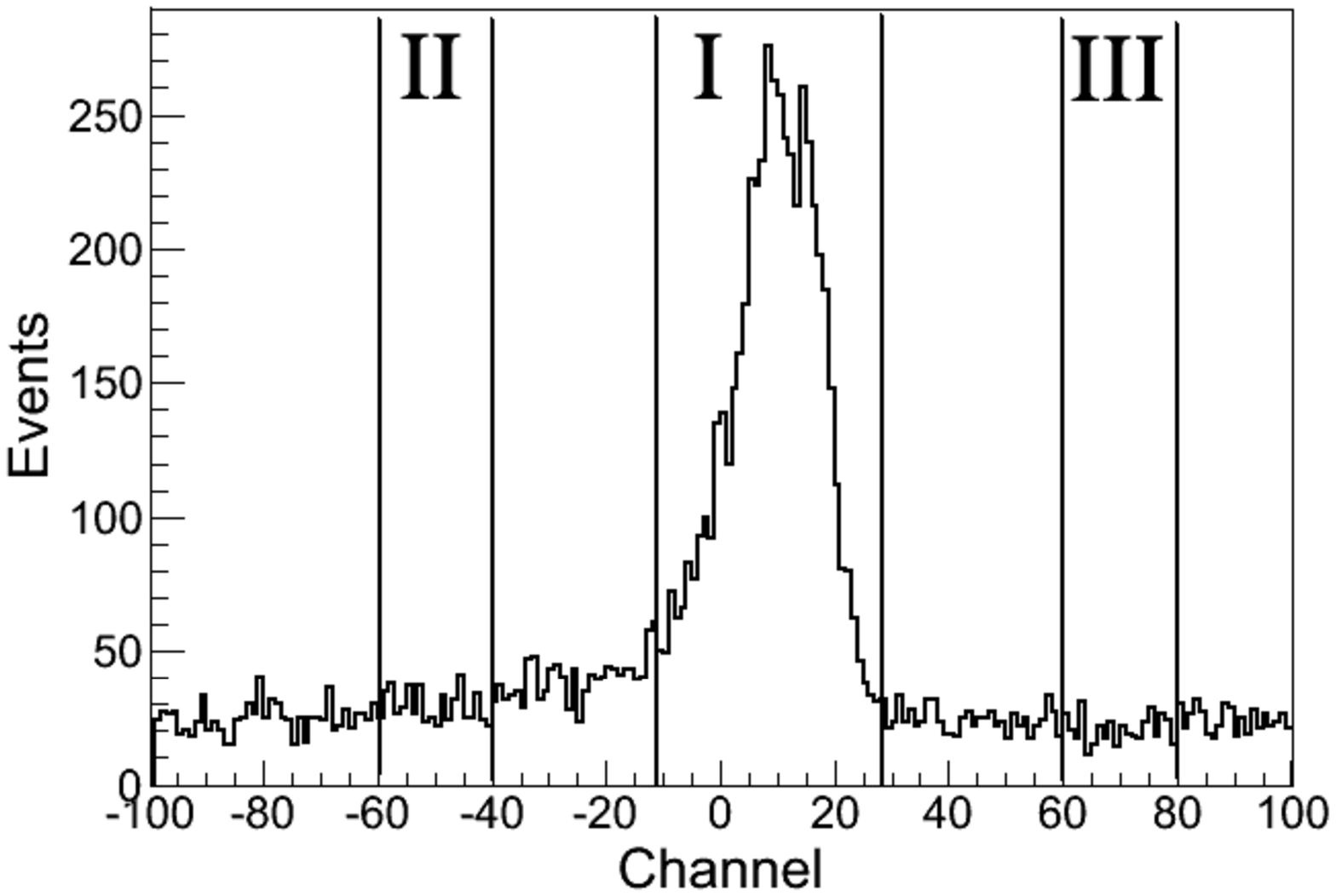}
\hfill
\caption{\footnotesize The time difference of   signals arrival from the deuteron and proton counters for $\theta_{c.m.} = 70^o$ in the c.m.s.}
\label{3}
\end{multicols}
\end{figure}
\par The amplitude distribution for proton counter by using these timing gates is shown in Fig. 7 A. The subtraction of the resulting spectra allows reducing  of the background (Fig. 7 B).
 \begin{figure}[h]
  \begin{center}
    \includegraphics[width=103mm]{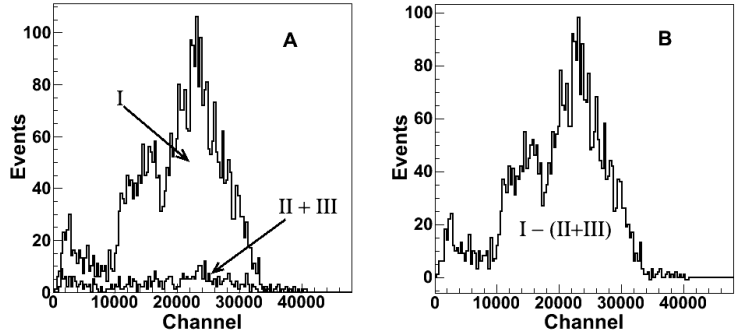}
  \end{center}
  \caption{\footnotesize The background subtraction procedure for the amplitude spectrum of proton counter for $\theta_{c.m.} = 70^o$ in the c.m.s..}
\label{fig:label}
\end{figure} 
\par The above analysis corresponds to the data obtained using CH$_2$-target. Analogous procedure was performed for the case with the C-target.
 \par The next stage is the CH$_2$-C subtraction procedure. The carbon background  subtraction normalization coefficient $k$  is deduced from the interval $a_{min}<a<a_{max}$, where $a$ - channels of CH$_2$- and C-amplitude distributions:  
\begin{eqnarray}
 k = \frac{N_{CH_2}|_{a_{min}<a<a_{max}}}{N_C|_{a_{min}<a<a_{max}}}.
\end{eqnarray} 
 Here $N_{CH_2}$ and $N_C$ - $CH_2$- and $C$-amplitude distributions integrals in $a$-interval within the window shown in Fig. 8 A by the solid lines. The carbon background can be then subtracted as: 
\begin{eqnarray}
 N_{dp} = N_{CH_2} - kN_C,
\end{eqnarray}
 where $N_{dp}$  is the resulting $dp$-elastic scattering distribution, $N_{CH_2}$ is the total $CH_2$-distribution, $kN_C$ is the normalized $C$-distribution within the window shown in Fig. 8 B by the dotted lines.
\par In Fig.8 A the CH$_2$-distribution is shown by the solid line. The normalized C-spectrum is shown by the dotted line. In Fig.8 B the result of subtraction is demonstrated. Such procedure was performed for proton amplitude spectra for each $\theta_{c.m.}$. 
 
\begin{figure}[h]
  \begin{center}
    \includegraphics[width=105mm]{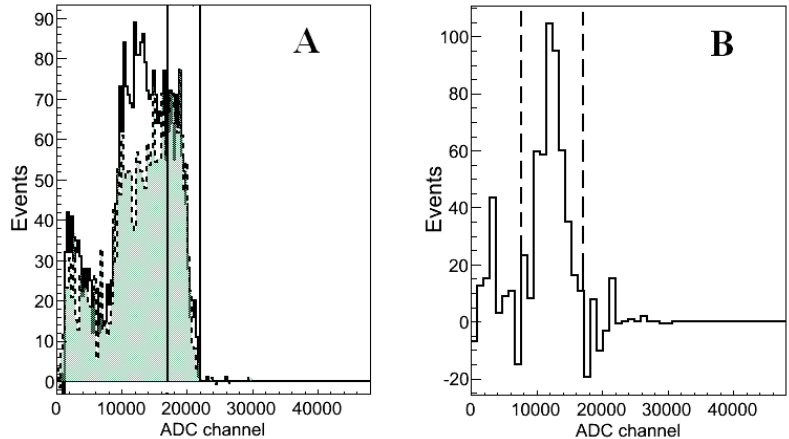}
  \end{center}
  \caption{\footnotesize The procedure of CH$_2$ - C subtraction for $\theta_{c.m.} = 70^o$ in the c.m.s. A - is the CH$_2$- and normalized C- distributions given by the solid and dotted histograms,respectively, vertical solid lines - is the interval of the normalization. B - is the result of $CH_2 -  C$ subtraction, vertical dashed lines are the gates indicating the domain of the $dp$-elastic scattering events.}
\label{fig:label}
\end{figure}


 \section{Results and discussions}

\vskip 5mm

\par The cross section was calculated using normalization to world data \cite{Bennet} at $\theta_{c.m.} = 70^o$ having the value $(d\sigma/d\Omega)|_{70^o} = 0.024 \pm 0.002$ mb/sr:
\begin{eqnarray}
\left(\frac{d\sigma}{d\Omega}\right)_{c.m.} = \frac{N_{dp}}{\Delta\Omega^{dp}_{lab}}\frac{C_{norm}}{N_M}J.
\end{eqnarray}
Here $N_{dp}$ - is the number of $dp$-elastic scattering events (after proper background subtraction), $\Delta\Omega^{dp}_{lab}$ - the effective angular span of proton detectors in the laboratory system, $N_M$ - is the number reconstructed events from one of the PP-counters, $C_{norm}$ - is the normalization coefficient, obtained  at 70$^o$ c.m.s. 
\par The transition from lab to the c.m. frame for differential cross section takes place by the transformation jacobian  $J$
which can be obtained by the kinematic calculations as:
\begin{eqnarray}
J = \frac{dcos\theta_{lab}}{dcos\theta_{c.m.}}.
\end{eqnarray} 
\par The effective angular span of proton detectors was calculated with the Pluto simulation package \cite{Pluto}. The value of the $d\Omega_{lab}$  decreases with the increasing of the $\theta_{c.m.}$. 
\par The statistical error associated with the background subtraction (eq. (2)) is roughly given by   $\delta N_{dp}^{stat} = \sqrt{N_{CH_2} + k^2N_C}$. The systematic error is due to normalization and $CH_2 - C$ subtraction procedure. The latter is defined as $\delta N_{dp}^{sys} = \delta kN_C$. The mean value of $\delta k \approx 5\%$. In general, the uncertainty of the normalization coefficient $C$ can be expressed from eq. 3 as:
\begin{eqnarray}
\delta C_{norm} = C\sqrt{\left(\frac{\delta \sigma}{\sigma}\right)^{2} + \left(\frac{\delta N_M}{N_M}\right)^{2} +  \left(\frac{\delta N_{dp}^{stat}}{N_{dp}}\right)^{2} + \left(\frac{\delta N_{dp}^{sys}}{N_{dp}}\right)^{2}}.
\end{eqnarray}
Here all values are defined for the $\theta_{c.m.} = 70^o$. $\delta \sigma/\sigma$ - is the  differential cross section relative error which equal about 8\% \cite{Bennet}. The value $\delta N_M/N_M$ is negligible. In particular, the uncertainty of  normalization  for the data for the $\theta_{c.m.} = 70^o$ is determined by the first and fourth term only. The  total systematic error varies in the range from 17\% to 40\%. 
\par The theoretical predictions for the differential cross section have been obtained in the relativistic multiple-scattering theory frame \cite{Ladygina}. In this model the reaction amplitude is defined by the corresponding transition operator. This operator obeys the Alt-Grassberger-Sandhas (AGS) equation \cite{Alt,Schmid}. After iteration these equations up to second-order term over NN $t$-matrix the reaction amplitude is defined as a sum of the three terms, which correspond to one-nucleon exchange ($ONE$), single scattering ($SS$) and double scattering ($DS$) reaction mechanisms. Since the ONE term gives a considerable contribution only at backward angles, this term was not included into consideration. Thus, the reaction amplitude is defined as a sum of two terms only. 
Diagrams for $SS$ and  $DS$ are presented in Fig. 9. All calculations were performed with the CD Bonn deuteron wave function \cite{Machleidt}.  The parameterization of the NN $t$-matrix was based on the use of the modern phase shift analysis \cite{shift} results.
\begin{figure}[h]
\begin{center}
\resizebox{0.35\textwidth}{!}{%
  \includegraphics{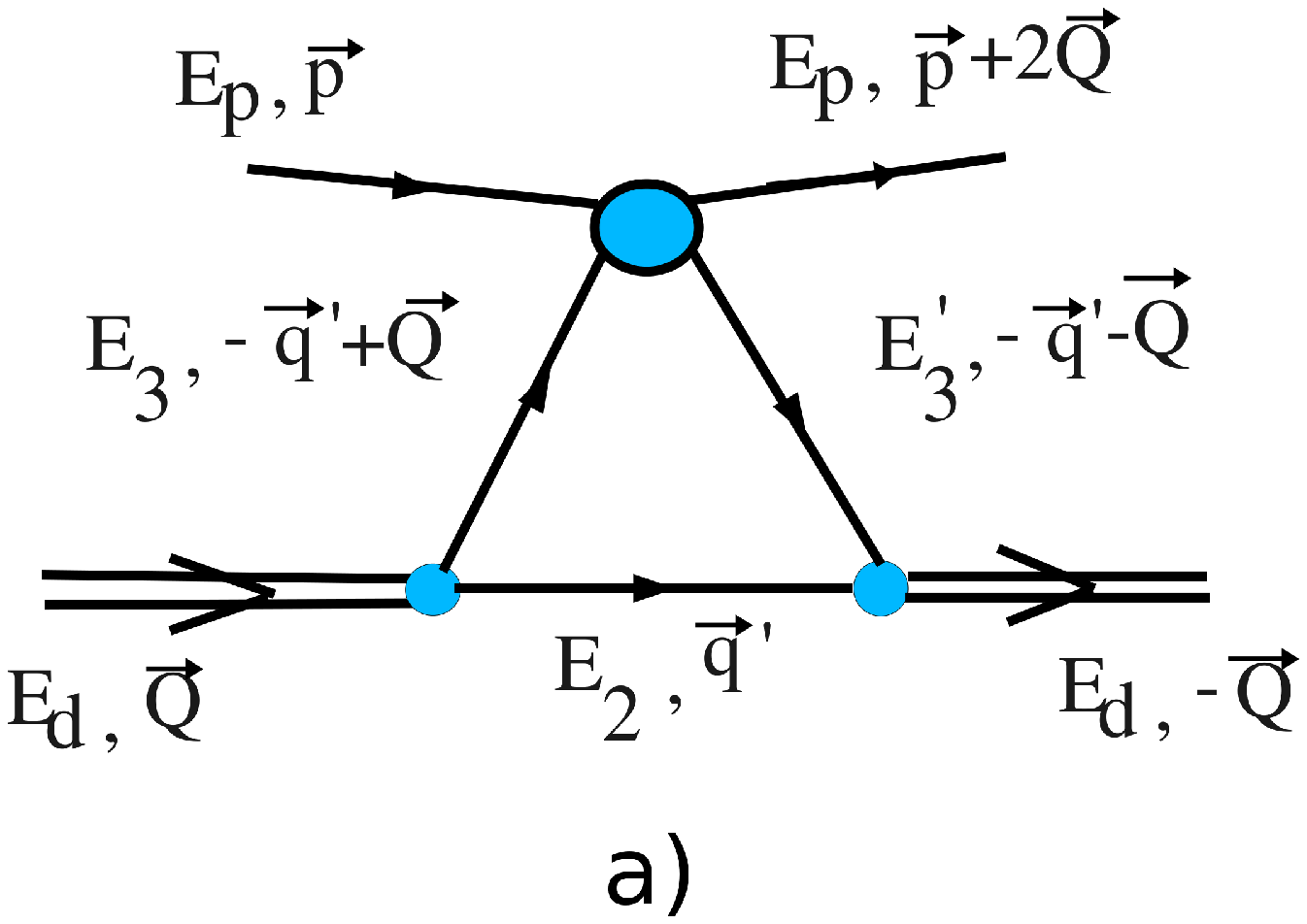}
}
\resizebox{0.40\textwidth}{!}{%
  \includegraphics{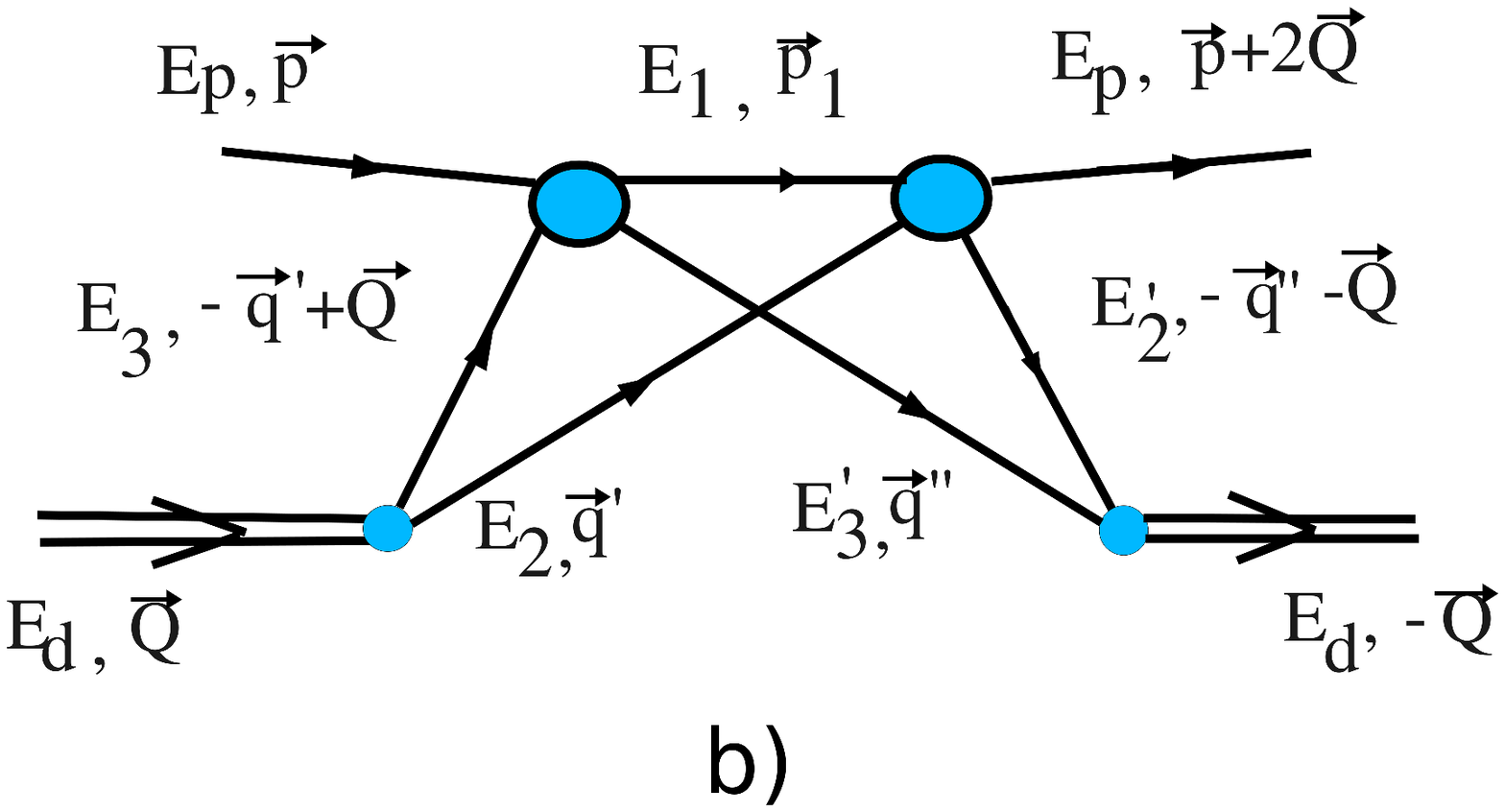}
}
\end{center}
\caption{\small{The diagrams taken into consideration for the calculations within relativistic multiple-scattering
model \cite{Ladygina}: a) - single scattering, b) - double scattering.}}
\label{fig:fig5}       
\end{figure}

\par In Fig.10 the data for differential cross section at 2 GeV are  compared with the world data and with the theoretical predictions. The new data are shown by the solid squares. The errors are the statistical only. The systematic errors are shown by the solid gray band. The data obtained earlier for forward angles \cite{Terekhin} are shown by the solid circles. The open triangles are world data from \cite{Bennet} obtained with a monochromatic protons beam at the Brookhaven Cosmotron by using a liquid-deuterium target.  
\begin{figure}[h]
  \begin{center}
    \includegraphics[width=11.5cm]{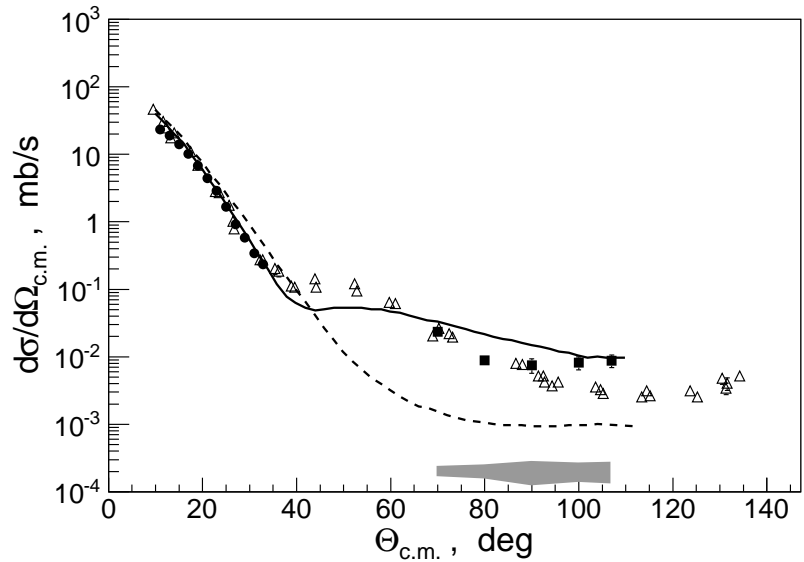}
  \end{center}
  \caption{\footnotesize The differential cross section for $dp$-elastic scattering at 1000 MeV/n. Squares - the results of this work, solid band - the systematic errors, circles - data from \cite{Terekhin}, triangles - data from \cite{Bennet}. The dashed and solid lines are the calculations without and with $DS$ term, respectively.}
\label{fig:label}
\end{figure}
The dashed and solid lines are the calculations without and with $DS$ term, respectively. One can see that the new data at $\theta_{c.m.} \leq 90^o$ are in good agreement with the world data. The discrepancy is increased at large angles, nevertheless, the data are in agreement with the errors which  increase with the angle increasing. On the other hand, there is better agreement with theoretical calculations taking into account DS at these angles. 
\par Fig. 10 shows that the single scattering mechanism does not reproduce  the experimental data at the scattering angles $\theta^*$ larger than $45^\circ$. The inclusion of the double scattering term in the calculations provides better agreement with the experimental results. However, some discrepancy remains. Probably, taking into account new reaction mechanisms like explicit $\Delta$-isobar excitation will improve the description of the data. 

\vskip 5mm

\textbf{\Large Conclusion}

\vskip 5mm

\par The procedure to obtain  differential cross section in $dp$-elastic scattering is shown. In the experiment the deuteron beam at energy 2 GeV, the CH$_2$- and C- targets were used. The data analysis was performed by using the CH$_2$ - C subtraction for amplitude spectra of proton detectors. 
\par The differential cross section data are obtained for the angular range of $70^\circ-107^\circ$ in the c.m.s. The results were compared with the data obtained early at the Synhrophasotron \cite{Terekhin} and the world data obtained at the Brookhaven Cosmotron \cite{Bennet}. The shape of the angular dependence of the relatively normalized data obtained at the Nuclotron agrees with the behaviour of the previously obtained data \cite{Bennet}. The discrepancy is increased at large angles, but nonetheless data are in agreement with achieved experimental accuracy.
\par The data are compared with the calculations performed within the framework of the relativistic multiple scattering theory \cite{Ladygina}. It is shown that taking into account the double scattering term improves the description of the obtained experimental results.

\begin{sloppypar}
The work has been supported in part by the RFBR grant 13-02-00101a.
\end{sloppypar}

\end{document}